\documentclass[showpacs,floatfix,preprint,nofootinbib]{revtex4-1}
\usepackage{graphicx,psfrag,amsmath,amssymb,amsfonts,latexsym,color,epsf,graphpap,dcolumn}
\usepackage{caption}
\usepackage{subcaption}

\begin{document}
\title{Casimir energy due to inhomogeneous thin plates}
\author{C.~D.~Fosco  and F.~D.~Mazzitelli}
\affiliation{Centro At\'omico Bariloche,  CONICET,
Comisi\'on Nacional de Energ\'\i a At\'omica, R8402AGP Bariloche, Argentina\\
and\\
Instituto Balseiro, Universidad Nacional de Cuyo, R8402AGP Bariloche, Argentina. }
\begin{abstract} 
We study the Casimir energy due to a quantum real scalar field coupled to
	two planar, infinite, zero-width, parallel mirrors with
	non-homogeneous properties. These properties are represented, in
	the model we use, by scalar functions defined on each mirror's
	plane. Using the Gelfand-Yaglom's theorem, we construct a
	Lifshitz-like formula for the Casimir energy of such a system. Then
	we use it to evaluate the energy perturbatively, for the case of
	almost constant scalar functions, and  also implementing a
	Derivative Expansion, under the assumption that the spatial
	dependence of the properties is sufficiently smooth. We point out
	that, in some particular cases, the Casimir interaction energy for
	non-planar perfect mirrors can be reproduced by inhomogeneities on
	planar mirrors.
\end{abstract}
\date{\today}
\maketitle
\section{Introduction}\label{sec:intro}
The evaluation of observables in the static Casimir effect presents many
interesting challenges, both from the point of view of designing
experiments capable of measuring its features with finer resolutions, as
well as regarding the calculations involved in the prediction of those
features \cite{books}.  In theoretical studies, the two main aspects
usually considered in a system exhibiting the Casimir effect are the
geometry of the mirrors and the properties of the media composing them. 

In some special cases, for example when the mirrors are assumed to satisfy
`perfect' boundary conditions~\footnote{Namely, boundary conditions which
do not involve any parameter or function; typically, Dirichlet or Neumann
for the case of a real scalar field, and perfect conductor conditions for the electromagnetic field.}, 
the ensuing simplifications allow one to study
geometrical aspects in a simpler way. The Derivative Expansion (DE), is an
approach that may be applied to the case of two smoothly curved surfaces,
such that at each point their local separation is smaller than the
curvature radii \cite{DE1,DE2}.  When the mirrors impose simple boundary
conditions, the DE predicts the {\em form\/} of the  corrections to the
Proximity Force Approximation (PFA), which is the leading term in the
expansion.  Indeed, the next to leading order  (NTLO) term is completely
fixed except for a single, global, dimensionless factor. That factor
depends exclusively on the kind of boundary condition used, and it may be
determined, once and for all, from a calculation which is perturbative in
the departure from planar mirrors, to the second order in that departure.
Note that the DE so obtained is not perturbative in the departure, yet it
can be determined, for example, from the knowledge that perturbative expansion.

The opposite, complementary case, would amount to systems with the simplest
possible, yet non trivial, geometry, like media having planar and parallel
(inter)faces.  Here, one can focus on the effects of the media
inhomogeneities on the detailed properties of the Casimir force, in this
case, its dependence on the distance between the mirrors. We are
particularly interested in situations where the planar media are
inhomogeneous at a macroscopic or mesoscopic level, that is, composed by
regular patches or periodic arrays of homogeneous media. On general
grounds, we expect  the Casimir energy to exhibit a rich dependence on the
inhomogeneities, similar to  those that appear when considering perfect
conductors with non trivial geometries, as trench arrays and other periodic
structures \cite{trenchs}.  Such rich dependence was measured in good agreement 
with calculations for sinusoidally corrugated surfaces \cite{demo}.

In this article, we consider the static Casimir effect for a special class
of system, consisting of a quantum real scalar field in the presence of two
zero-width, parallel and infinite plates. The latter are not assumed to be
homogeneous; rather, each plate will be characterized, regarding its
properties, by a real scalar potential defined on the plane.  This can be
considered as a toy model for the interaction of the electromagnetic field
with a thin mirror characterized by position-dependent electromagnetic
properties, and generalizes previous works for scalar fields in the
presence of homogeneous thin plates, named $\delta$-potentials
\cite{deltap}. Under those assumptions about the system, and using the
Gelfand-Yaglom's (GY) theorem \cite{GY}, we shall obtain a Lifshitz-like
formula \cite{Lifshitz}, which yields the Casimir energy as a functional
having as arguments the `scalar potentials', i.e., the two functions which
characterize the plates \cite{CcapaTtira}.
We then use this generalized Lifshitz-formula to obtain more explicit
expressions, making further assumptions about the potentials. We firstly
consider the case of scalar potentials describing small departures from
homogeneity, implementing a perturbative approach Casimir. 
We compare and relate the results to those
corresponding to two Dirichlet mirrors with non planar geometries.   On the other
hand, if  the space-dependence of the properties is
smooth, a DE for the energy becomes a useful tool. We use the perturbative
results to obtain the usual PFA and its NTLO for this particular system.  

This paper is organized as follows: in Sect.~\ref{sec:lif}, we derive the
would-be Lifshitz formula for the system. Then, in Sect.~\ref{sec:exp}, we
present the expansion up to the second order in the departures from homogeneity,
and discuss some particular limiting cases. In
Sect.~\ref{sec:de} we apply the previous result to the calculation of the
DE to the second order in derivatives. Section~\ref{sec:conc}  contains our
conclusions.

\section{Lifshitz formula for non homogeneous plates}\label{sec:lif}
We derive here an expression for the vacuum energy corresponding to a
quantum real scalar field $\varphi(x)$ in $3+1$ dimensions, coupled to two
non-homogeneous zero-width, infinite, parallel plates. 
It may be regarded as a Lifshitz formula but for zero-width inhomogeneous
plates, or even as a special case of the TGTG formula \cite{TGTG}. 

The kind of model we deal with may be defined in terms of its Euclidean
action ${\mathcal S}(\varphi)$, which is assumed to be of the form:
\begin{equation}\label{ea:defsphi}
{\mathcal S}(\varphi) \;=\; \frac{1}{2} \, \int d^4x \, 
\left[ \big( \partial \varphi (x) \big)^2 + V({\mathbf x}) \big( \varphi(x)
\big)^2 \right]  \;,
\end{equation}
with spacetime coordinates $x = (x_0,x_1,x_2,x_3)$, 
and spatial coordinates ${\mathbf x}=(x_1,x_2,x_3)$.  The `potential' $V$
accounts for the presence of two plates, denoted by $R$ and $L$, and
located on the planes $x_3=\frac{l}{2}$ and $x_3=-\frac{l}{2}$,
respectively. Specifically, that function is assumed to be of the form:
\begin{equation}\label{potentials}
	V({\mathbf x}) \;=\; \delta(x_3 -\frac{l}{2}) \,
	V_R({\mathbf x_\shortparallel}) \,+\, \delta(x_3 +\frac{l}{2}) \,
	V_L({\mathbf x_\shortparallel}) \;,
\end{equation}
where ${\mathbf x_\shortparallel}=(x_1,x_2)$ are coordinates on the
mirrors. We could also include a bulk potential
$V_{bulk}(x_3,{\mathbf x_\shortparallel})$ between mirrors. This is a very
interesting problem that raises mathematical and physical questions about
the proper derivation of finite Casimir forces \cite{FulMil}. We plan to
address this issue in a forthcoming paper. It is worth to stress that the
potentials $V_{R,L}$ on the plates mimic inhomogeneities of the mirrors,
that could be formed by a juxtaposition of homogeneous materials. Other
kind of situations, like  frequency-dependent electromagnetic properties,
would require, to be modelled, the presence of terms in the action which
are nonlocal in time.   

In the approach that we shall follow to derive the generalized Lifshitz
formula, the vacuum energy will be obtained from the functional integral:
\begin{equation}
{\mathcal Z} \;=\; \int {\mathcal D}\varphi \; e^{- {\mathcal S}(\varphi)}
\;.
\end{equation} 
Indeed, assuming the system to be defined  within a `time interval':
$-\frac{T}{2} < \tau < \frac{T}{2}$ and considering the
$T \to \infty$ limit, since the  potential has been assumed to be
time-independent, the leading behaviour of the effective action $\Gamma$ is 
$\Gamma \, \sim \, T \times E$, $E$ being the vacuum energy. Thus, we may
relate $E$ to a functional determinant, by using the formal result for 
$\Gamma$ which, ignoring irrelevant factors, is given by:
\begin{equation}\label{eq:defdet}
	e^{-\Gamma} \;=\;  \big[\det(-\partial^2 + V
	\big)\big]^{-\frac{1}{2}} \;,
\end{equation}
with $\partial^2 \equiv \partial_\mu \partial_\mu$,  $\mu = 0,1,\ldots,d$.

Therefore
\begin{equation}\label{eq:gen}
E \,=\, \frac{1}{2} \, \int_{-\infty}^{+\infty} \frac{dk_0}{2\pi} \; 
\log \big(\det{\mathbb K}\big) \;=\;
\frac{1}{2} \, \int_{-\infty}^{+\infty} \frac{dk_0}{2\pi} \; 
{\rm Tr} \big(\log{\mathbb K}\big) \;,
\end{equation}
where
\begin{equation}\label{eq:defkgen}
	{\mathbb K}\;=\; -\nabla_{\mathbf x}^2 + k_0^2 \,+\, V({\mathbf x})\;,
\end{equation}
and the determinant and trace are understood to be in the functional sense,
for operators acting on functions of ${\mathbf x}$.

Our strategy to derive the generalized Lifshitz formula consists of
`dimensionally reducing' the determinant appearing in (\ref{eq:gen}). By
this, we mean transforming it into one where the $x_3$ coordinate has been
dealt with. Equivalently, at the level of the functional trace, it amounts to
tracing out the $x_3$ coordinate, writing the result in terms of
operators now acting on functions of ${\mathbf x}_\shortparallel \equiv
(x_1, x_2)$ (rather than ${\mathbf x}_\shortparallel \equiv (x_1, x_2)$).
That can be done by a suitable application of GY theorem, as follows.  We first
assume the system to be defined  within a large `box' along the $x_3$
direction, namely: $-\frac{L}{2} < x_3 < \frac{L}{2}$ where $L >> l$, and
we shall take the $L\to \infty$ limit at the end of the calculation. 
To proceed, we will need to assume  some boundary conditions on $x_3=\pm
\frac{L}{2}$. We shall, for the sake of simplicity, use Dirichlet
conditions~\footnote{The specific choice of boundary conditions  has no bearing on
the results, since we will take the $L \to \infty$ at the end of the
calculation.}. 

We first rewrite the operator ${\mathbb K}$ introduced above in a way which
makes it simpler to perform the dimensional reduction:  
\begin{equation}\label{eq:ksplit}
	{\mathbb K} \;=\; -\partial_3^2 + {\mathbb H}(x_3) \;,
\end{equation}
where:
\begin{equation}
{\mathbb H}(x_3) \;\equiv\; - \nabla_\shortparallel^2 \,+\,
	V_{x_3}({\mathbf x}_\shortparallel) \, +\, k_0^2 \;,
\end{equation}
with 
\begin{equation}
V_{x_3}({\mathbf x}_\shortparallel) \;\equiv \;V({\mathbf x}) \;\;,\;\;\;
\nabla_\shortparallel \;\equiv\; \nabla_{{\mathbf x}_\shortparallel} \;.
\end{equation}	
The reason for this seemingly formal splitting is that $x_3$ is
the variable about which the functional determinant of ${\mathbb K}$ will
be reduced.  To each value of $x_3$, we may associate an operator kernel along the
${\mathbf x}_\shortparallel$ coordinates; indeed: 
\begin{align}
\langle {\mathbf x}_\shortparallel | {\mathbb K} | {\mathbf x}'_\shortparallel \rangle 
&\;=\; K_{x_3}({\mathbf x}_\shortparallel , {\mathbf x}'_\shortparallel) \nonumber\\
&\;=\; -  \delta({\mathbf x}_\shortparallel - {\mathbf x}'_{\shortparallel}) \partial_3^2
\,+\, H_{x_3}({\mathbf x}_\shortparallel , {\mathbf x}'_{\shortparallel}) \;,
\end{align}
where
\begin{align}
& H_{x_3}({\mathbf x}_\shortparallel , {\mathbf x}'_{\shortparallel}) \,=\, 
\langle {\mathbf x}_\shortparallel | {\mathbb H}(x_3) | {\mathbf
	x}'_\shortparallel \rangle \nonumber\\
&\;=\; - \nabla_{{\mathbf x}_\shortparallel}^2 \delta({\mathbf
	x}_\shortparallel - {\mathbf x}'_{\shortparallel})  \,+\, 
	[ V_{x_3}({\mathbf x}_\shortparallel) + k_0^2 ]  \, \delta({\mathbf
	x}_\shortparallel - {\mathbf x}'_{\shortparallel}) \;.
\end{align}

The corresponding result for the determinant is usually presented in terms of its ratio
with a `reference' operator ${\mathbb K}_0$, which in our case we assume to
correspond to a vanishing potential. Using this ratio instead of just ${\mathbb K}$ in
(\ref{eq:gen}) is indeed convenient, since it then produces the energy of
the system taking as a reference the energy of the vacuum in the absence of
the plates.

The ratio between determinants may be put in terms of two
determinants of dimensionally reduced operators~\cite{Kleinert:2004ev}:
\begin{equation}\label{eq:gyf}
\frac{\det{\mathbb K}}{\det{\mathbb K}_0}\;=\;
\frac{\det \psi(\frac{L_d}{2})}{\det\psi_0(\frac{L_d}{2})} \;,
\end{equation}
where  $\psi(x_d)$ and $\psi_0(x_d)$ are {\em operatorial\/} solutions to
the homogeneous equations
\begin{equation}\label{eq:hom1}
	{\mathbb K} \psi(x_d) \;=\; 0 \;\;,\;\;\;
	{\mathbb K}_0 \psi_0(x_d) \;=\; 0 \;,
\end{equation}
The matrix elements of these solutions may naturally be denoted by:
\begin{equation}
\langle {\mathbf x}_\shortparallel | \psi(x_3) | {\mathbf
	x}'_\shortparallel \rangle \;=\;
\psi_{x_3}( {\mathbf x}_\shortparallel , {\mathbf x}'_\shortparallel) \;,
\end{equation}
(and analogously for $\psi_0$). The initial conditions on the solutions to
the homogeneous equations are
\begin{align}
	& \big[ \psi_{x_3}( {\mathbf x}_\shortparallel , {\mathbf
	x}'_\shortparallel)
\big]\Big|_{x_3 = - \frac{L}{2}} \;=\; 0 \nonumber\\
& \big[ \frac{\partial \psi_{x_3}}{\partial x_3}({\mathbf x}_\shortparallel ,
{\mathbf x}'_\shortparallel )
\big]\Big|_{x_3 = - \frac{L_3}{2}} \;=\;  
\delta( {\mathbf x}_\shortparallel - {\mathbf x}'_\shortparallel) \;.
\end{align}

We need a more explicit solution to Eq.(\ref{eq:hom1}); that can be done by
first converting it to a first order system, by introducing
\begin{equation}
	\Psi(x_3)\;\equiv\; 
	\left( 
		\begin{array}{c}
			\psi (x_3)\\
			\frac{\partial \psi}{\partial x_3}(x_3) 	
		\end{array}
	\right) \;\equiv\;
	\left( 
		\begin{array}{c}
			\Psi_1 (x_3)\\
			\Psi_2 (x_3)
		\end{array}
	\right) \;,
\end{equation}
which renders the original second-order equation into 
\begin{equation}\label{eq:hom2}
	\frac{\partial \Psi}{\partial x_3}(x_3) \;=\; {\mathcal H}(x_3)
	\Psi(x_3) \;.
\end{equation}
with
\begin{equation}
{\mathcal H}(x_3) \;=\;	
	\left( 
		\begin{array}{cc}
			0 &  {\mathbb I}		\\
			{\mathbb H}(x_3) & 0 
		\end{array}
	\right) \;,
\end{equation}
where ${\mathbb I}$ denotes the identity operator. Note that
Eq.\eqref{eq:hom2} is a Schroedinger-like equation in which the coordinate
$x_3$ plays the role of time, and  ${\mathcal H}$ the role of a
`Hamiltonian'.  In terms of the evolution operator ${\mathcal U}$, we then have
the solution:
\begin{equation}
	\Psi(x_3) \;=\; {\mathcal U}(x_3,-\frac{L}{2}) \Psi(-\frac{L}{2}) 
\end{equation}	
with
\begin{equation}\label{eq:defu}
	{\mathcal U}(x''_3,x'_3) \,\equiv\,
	{\mathcal P} \exp\Big[ \int_{x'_3}^{x''_3} dy_3
	{\mathcal H}(y_3) \Big] \;\;\;\; (x''_3 \geq x'_3) \;,
\end{equation}
with the path-ordering operator ${\mathcal P}$, which acts 
in the same way as the time-ordering operator, but with $x_3$ 
playing the role of the time.

Equipped with the solution for $\Psi$ just presented, we note that
\begin{equation}\label{eq:gyf1}
\left( \begin{array}{c}
\psi(\frac{L}{2}) \\
     0
\end{array} 
\right) \,=\, {\mathcal U}(\frac{L}{2},-\frac{L}{2}) \left( \begin{array}{c}
    0 \\
{\mathbb I}
\end{array} 
\right)\;,
\end{equation}
or, using indices $A$ and $B$, which can assume the values $1$ or $2$, to 
distinguish the $4$ (operatorial) blocks in ${\mathcal U}(\frac{L}{2},-\frac{L}{2})$:
\begin{equation}\label{eq:mblocks}
{\mathcal U}(\frac{L}{2},-\frac{L}{2}) \;\equiv\; 
\left( \begin{array}{cc}
{\mathcal U}_{11}(\frac{L}{2},-\frac{L}{2})  & {\mathcal U}_{12}(\frac{L}{2},-\frac{L}{2})  \\
{\mathcal U}_{21}(\frac{L}{2},-\frac{L}{2})  & {\mathcal U}_{22}(\frac{L}{2},-\frac{L}{2})  
\end{array} 
\right)\;,
\end{equation}
we see that 
\begin{equation}
\frac{\det{\mathbb K}}{\det{\mathbb K}_0}\;=\; \frac{\det{\mathcal
U}_{12}(\frac{L}{2},-\frac{L}{2})}{\det{\mathcal U}_{12}^{(0)}(\frac{L}{2},-\frac{L}{2})}\, .
\end{equation}

This is a more explicit form of the reduction, where we still need to take
the $L\to \infty$ limit. To do this we note that, since $V$ vanishes for
$|x_3| > \frac{l}{2}$, we may write:
\begin{equation}
	{\mathcal U}(\frac{L}{2},-\frac{L}{2}) \;=\;\lim_{\epsilon \to 0^+}
	\Big[ {\mathcal
	U}^{(0)}(\frac{L}{2},\frac{l}{2} +\epsilon) \,
	{\mathcal U}(\frac{l}{2}+\epsilon,-\frac{l}{2} - \epsilon) \, 
	{\mathcal U}^{(0)}(-\frac{l}{2}-\epsilon,-\frac{L}{2}) \Big]
\end{equation}
where the $\epsilon \to 0^+$ limit has been introduced, since the potential is
discontinuous at $\pm \frac{l}{2}$. 

The explicit form of ${\mathcal U}^{(0)}(x'_3,x''_3)$ may be found by
exponentiation, because it corresponds to an $x_3$-independent potential:
\begin{align}\label{eq:umat1}
	{\mathcal U}(x'_3,x''_3) \;=\; e^{(x'_3-x''_3) {\mathcal H}_0} \;& =\;
	\frac{\sinh((x'_3 -x''_3) \sqrt{{\mathbb H}_0})}{\sqrt{{\mathbb
	H}_0}} \, 
\left( \begin{array}{cc}
	 0 &  {\mathbb I}		\\
	{\mathbb H}_0 & 0 
	\end{array}
\right) \nonumber\\
	\,&+\,\cosh((x'_3-x''_3) \sqrt{{\mathbb H}_0}) \, 
\left( \begin{array}{cc}
		{\mathbb I}	& 0	\\
		0  & {\mathbb I}  
		\end{array}
\right) \;.
\end{align}

Using this result, a rather lengthy but otherwise straightforward
calculation shows that, when $L\to\infty$, the ratio between determinants
may be written in terms of $\mu_{AB}$, a shorthand notation for the matrix
elements
\begin{equation}
	\mu_{AB}\, =\,\lim_{\epsilon \to 0^+}  \Big[{\mathcal
	U}_{AB}(\frac{l}{2}+\epsilon,-\frac{l}{2}-\epsilon) \Big] 
\end{equation}
as follows:
\begin{equation}\label{eq:gyfa}
\frac{\det{\mathbb K}}{\det{\mathbb K}_0}\;=\;
	\det\left[ 
	\frac{1}{2} e^{-\frac{l}{2} \sqrt{{\mathbb H}_0}}
	\big(
	\mu_{11} + \mu_{12}  \sqrt{{\mathbb H}_0} +
	\frac{1}{\sqrt{{\mathbb H}_0}} \mu_{21} + 
	\frac{1}{\sqrt{{\mathbb H}_0}} \mu_{22} 
	\sqrt{{\mathbb H}_0}
	\big)
	 e^{-\frac{l}{2} \sqrt{{\mathbb H}_0}}
	\right]
\end{equation}

The vacuum energy, referred to the vacuum in the absence of the plates, is
then given by
\begin{align}\label{eq:ess}
E \;=\; \frac{1}{2} \,\int_{-\infty}^{+\infty} \frac{dk_0}{2\pi}
	& \Big\{ \log \det \big[\frac{1}{2} (\mu_{11} + \mu_{12}  \sqrt{{\mathbb H}_0} +
	\frac{1}{\sqrt{{\mathbb H}_0}} \mu_{21} + 
	\frac{1}{\sqrt{{\mathbb H}_0}} \mu_{22} 
	\sqrt{{\mathbb H}_0} )
	\big]  \nonumber\\
	& + \log \det e^{- l \sqrt{{\mathbb H}_0}} \Big\}
\;.
\end{align}
In the previous expression, no use has been made yet of the precise form of
the potential in the region occupied by the plates: this will determine the
objects we have denoted by $\mu_{AB}$. Dealing with the effect of the 
$\delta$ functions, one can derive the explicit form:
\begin{equation}
	\left(
	\begin{array}{cc}
		\mu_{11} & \mu_{12} \\
		\mu_{21} & \mu_{22}
	\end{array}
	\right) 
	\;=\; 
	\left( 
	\begin{array}{cc}
		1 & 0 \\
		V_R & 0
	\end{array}
        \right) 
	\Big[
		\frac{\sinh(l \sqrt{{\mathbb H}_0})}{\sqrt{{\mathbb H}_0}} \, 
\left( \begin{array}{cc}
	 0 &  {\mathbb I}\\
	{\mathbb H} & 0 
	\end{array}
\right) 
	\,+\,\cosh( l \sqrt{{\mathbb H}_0}) \, 
	\Big]
	\left( 
	\begin{array}{cc}
		1 & 0 \\
		V_L & 0
	\end{array}
	\right) \;,
\end{equation}
which, when inserted in (\ref{eq:ess}), yields for the energy a result with
the structure:
\begin{equation}
	E \;=\; E_L \,+\,E_R \,+\, E_I
\end{equation}
with $E_L$ ($E_R$) denoting the self-energy of the $L$ ($R$) plate:
\begin{equation}
	E_{L,R} \;=\; \frac{1}{2} \,\int_{-\infty}^{+\infty} \frac{dk_0}{2\pi}
	{\rm Tr}  \log \Big[ {\mathbb I}  \,+\, \frac{1}{2\sqrt{{\mathbb
	H}_0}} \, V_{L,R}({\mathbf x}_\shortparallel) \Big] \;,
\end{equation}
where, we recall, ${\mathbb H}_0 = - {\mathbf\nabla}_\parallel^2 + k_0^2$,
and $E_I$ is the interaction energy, responsible for the Casimir force
between mirrors. It is given by a generalized form of Lifshitz formula:
\begin{equation}\label{eq:lifshitz}
E_I \;=\; \frac{1}{2} \,\int_{-\infty}^{+\infty} \frac{dk_0}{2\pi}
{\rm Tr}  \log \Big[ {\mathbb I}  \,-\, e^{- l \sqrt{{\mathbb H}_0}} \,
r_R \, e^{- l \sqrt{{\mathbb H}_0}} \, r_L \, \Big] \;,
\end{equation}
with the operatorial `reflection coefficients':
\begin{equation}\label{eq:ref}
	r_{L,R} \,=\, ( 2\sqrt{{\mathbb H}_0} +
	V_{L,R})^{-1} \,V_{L,R}\;. 
\end{equation}
Note that, except when the potentials are uniform, i.e., independent of
${\mathbf x}_\shortparallel$, these reflection coefficients involve non
commuting objects in their definitions, since ${\mathbb H}_0$ depends on
the planar Laplacian operator. That makes it difficult to obtain more
explicit expressions, unless extra assumptions are made.

\section{Expansion up to the second order in the departure from constant
potentials}\label{sec:exp}
Let us now expand Eq.(\ref{eq:lifshitz}), the general expression for the
interaction part of the vacuum energy, up to the second order in the
departure from constant functions:
\begin{equation}
	V_L({\mathbf x}_\shortparallel) \,=\, v_L \, + \, \eta_L({\mathbf
	x}_\shortparallel) \;\;,\;\;\;\;
	V_R({\mathbf x}_\shortparallel) \,=\, v_R \, + \, \eta_R({\mathbf
	x}_\shortparallel) \;, 
\end{equation}
where $v_L$ and $v_R$ are constants. We assume they are chosen in such a
way that the spatial average of each departure vanish.
The expansion for the energy is then:
\begin{equation}
	E_I \;=\; E^{(0)} \,+\, E^{(1)}\,+\, E^{(2)} \,+\ldots
\end{equation}
where: 
\begin{equation}\label{E(0)}
	\frac{E^{(0)}}{L^2} \,=\, \frac{1}{32 \pi^2 l^3} \int_0^\infty d\rho
	\rho^2 \,\log\big[ 1 - \sigma_L(\rho) \sigma_R(\rho) e^{-\rho} \big] \;,  
\end{equation}
where $L^2$ is the area of the plates, and we have introduced
\begin{equation}
	\sigma_{L,R}(\rho) \;\equiv\; \frac{x_{L,R}}{\rho + x_{L,R}} 
	\;,\;\;\;
	x_{L,R} \;\equiv\;  l \, v_{L,R} \;.
\end{equation}
Note that the Dirichlet limit is obtained as $\sigma_{L,R}(\rho)\to 1$.
Eq.\eqref{E(0)} is easily obtained from Eq.\eqref{eq:lifshitz} by setting
$V_{L,R}=v_{L,R}$ in the reflection coefficients of Eq.\eqref{eq:ref}.

The first-order term, on the other hand, vanishes. Indeed, it has the form:
\begin{equation}\label{eq:e1}
	E^{(1)} \;=\; \int_{{\mathbf x}_\shortparallel} \,\Big[
	\sum_\alpha \frac{\delta E}{\delta \eta_\alpha({\mathbf
	x}_\shortparallel)}\Big|_{\eta_L\equiv 0,\,\eta_R \equiv 0} \, 
	\eta_\alpha({\mathbf x}_\shortparallel) \Big] \;. 
\end{equation}
We have adopted above some conventions that we shall continue to use: a
shorthand notation for the integration over ${\mathbf x}_\shortparallel$,
and indices from the beginning of the Greek alphabet to denote
each one of the two mirrors, namely, $\alpha = L, R$.

Now, the functional derivatives in (\ref{eq:e1}) are evaluated for
(simultaneously) vanishing $\eta_L$ and $\eta_R$, and are
thus ${\mathbf x}_\shortparallel$-independent. Therefore,
\begin{equation}
E^{(1)} \;=\; \sum_{\alpha=L,R} 
\;\frac{\delta E}{\delta \eta_\alpha }\Big|_{\eta_L\equiv 0, \eta_R
	\equiv 0} \, \int_{{\mathbf x}_\shortparallel} \,
\eta_\alpha({\mathbf x}_\shortparallel) \,=\,0 \;. 
\end{equation}

The second-order term, $E^{(2)}$, is a quadratic form in the
departures,
\begin{equation}
E^{(2)} \;=\; \frac{1}{2} \, \sum_{\alpha, \beta} 
\int_{{\mathbf x}_\shortparallel,{\mathbf x}'_\shortparallel} 
\,
\eta_\alpha({\mathbf x}_\shortparallel) \; \gamma_{\alpha\beta}({\mathbf
	x}_\shortparallel - {\mathbf x'}_\shortparallel)
	\;\eta_\beta({\mathbf x}'_\shortparallel) \;,
\end{equation}
where the $\gamma_{\alpha\beta}$, being second functional derivatives at
vanishing departures, can only depend on the differences between the
arguments. Moreover, one can also show that the depend just on the
modulus of ${\mathbf x}_\shortparallel - {\mathbf x}'_\shortparallel$,
since for constant potentials the plates are homogeneous and isotropic.

It is convenient to use, in what follows, Fourier transforms.
We use a tilde on the Fourier transformed version of an object, 
\begin{equation}
	\eta({\mathbf x}_\shortparallel) \,=\, \int \frac{d^2{\mathbf
	k}_\shortparallel}{(2\pi)^2} \; e^{i {\mathbf k}_\shortparallel
	\cdot {\mathbf x}_\shortparallel} \; \tilde{\eta}({\mathbf
	k}_\shortparallel) \;,
\end{equation}
and analogous conventions for Fourier transformation along the three
coordinates $x_0, x_1, x_2$ ($\equiv x_\shortparallel$), on the space-time
of the mirrors. As a technical remark: it is convenient, since the
expressions become more symmetrical, to use space-time dependent
departures, and to regard them as time-independent at the end of the
calculation.

Thus,
\begin{equation}
E^{(2)} \;=\; \frac{1}{2} \, \sum_{\alpha, \beta} \int\frac{d^2{\mathbf
k}_\shortparallel}{(2\pi)^2}  \; 
\tilde{\eta}_\alpha(-{\mathbf k}_\shortparallel) \;
	\tilde{\gamma}_{\alpha\beta}(|{\mathbf
	k}_\shortparallel|)
	\;\tilde{\eta}_\beta({\mathbf k }_\shortparallel) \;,
\end{equation}
where, having in mind the static limit mentioned above,
\begin{equation}
\tilde{\gamma}_{\alpha\beta}(|{\mathbf k}_\shortparallel|)
	\;=\; \lim_{k_0 \to 0} \Big[
	\tilde{\gamma}_{\alpha\beta}(k_\shortparallel) \Big] \;.
\end{equation}
To compute the $\tilde{\gamma}_{\alpha\beta}$, we need to expand the reflection coefficients
in Eq.\eqref{eq:ref} up to second order in $\eta_{L,R}\,  ,$ then insert the result in 
Eq.\eqref{eq:lifshitz}, and keep the quadratic terms. The expansion of the
reflection coefficients, which has of course the same form for each plate,
may be written as follows (we only write the one for the $L$ plate, the
expansion for the other plate is obtained by replacing $L \to R$ below):
\begin{equation}
	r_L \;=\; r_L^{(0)} \,+\, r_L^{(1)} \,+\, r_L^{(2)} \,+\, \ldots 
\end{equation}
where:
\begin{align}
r_L^{(0)} &=\; ( 2\sqrt{{\mathbb H}_0} + v_L)^{-1} \,v_L  \;,\nonumber\\
r_L^{(1)} &=\; \frac{1}{v_L^2} \, r_L^{(0)} \,  \eta_L \,  r_L^{(0)}\, 2
	\sqrt{{\mathbb H}_0} \;, \nonumber\\
r_L^{(2)} &=\; - \frac{1}{v_L^3} \, r_L^{(0)} \, \eta_L \,  r_L^{(0)} \eta_L \, r_L^{(0)}
	\, 2 \sqrt{{\mathbb H}_0} \;. 
\end{align}
The form of the order $n$ term, for any $n \geq 1$, is in fact:
\begin{equation}
r_L^{(n)} \;=\; (-1)^{n-1} \, \frac{1}{v_L^{n+1}} \, \big[
r_L^{(0)} \, \eta_L \big]^n \,  r_L^{(0)} \, 2 \sqrt{{\mathbb H}_0} \;. 
\end{equation}

Inserting the previous expansion for each one of the plates to the second
order, and expanding the logarithm accordingly, we may write down the
different components of $\tilde{\gamma}_{\alpha\beta}$ more explicitly. The
first ones we write down involve a departure in just one of the mirrors, to
the second order:
\begin{align}\label{eq:gll}
	&\tilde{\gamma}_{LL}(k_\shortparallel) \;=\; \frac{1}{2^3 x_L^3 l} \, \int
	\frac{d^3p_\shortparallel}{(2\pi)^3} \, 
	\frac{ |p_\shortparallel| \sigma_R(p_\shortparallel) 
	\sigma_L^2(p_\shortparallel) \sigma_L(p_\shortparallel + 2 l
	k_\shortparallel)}{e^{|p_\shortparallel|} -
	\sigma_L(p_\shortparallel) \sigma_R(p_\shortparallel)} \nonumber\\ 
	&-\; \frac{1}{2^4 x_L^4 l} \, \int
	\frac{d^3p_\shortparallel}{(2\pi)^3} \; 
	\frac{|p_\shortparallel| \sigma_R(p_\shortparallel) 
	\sigma_L^2(p_\shortparallel)}{e^{|p_\shortparallel|} -
	\sigma_L(p_\shortparallel) \sigma_R(p_\shortparallel)} \,
\frac{|p_\shortparallel + 2 l k_\shortparallel|
	\sigma_R(p_\shortparallel + 2 l k_\shortparallel)
	\sigma_L^2(p_\shortparallel + 2 l
	k_\shortparallel)}{e^{|p_\shortparallel + 2 l k_\shortparallel|} -
	\sigma_L(p_\shortparallel + 2 l k_\shortparallel)
	\sigma_R(p_\shortparallel + 2 l k_\shortparallel)} \;,
\end{align}
\begin{align}\label{eq:grr}
	&\tilde{\gamma}_{RR}(k_\shortparallel) \;=\; \frac{1}{2^3 x_R^3 l} \, \int
	\frac{d^3p_\shortparallel}{(2\pi)^3} \, 
	\frac{ |p_\shortparallel| \sigma_L(p_\shortparallel) 
	\sigma_R^2(p_\shortparallel) \sigma_R(p_\shortparallel + 2 l
	k_\shortparallel)}{e^{|p_\shortparallel|} -
	\sigma_R(p_\shortparallel) \sigma_L(p_\shortparallel)} \nonumber\\ 
	&-\; \frac{1}{2^4 x_R^4 l} \, \int
	\frac{d^3p_\shortparallel}{(2\pi)^3} \; 
	\frac{|p_\shortparallel| \sigma_L(p_\shortparallel) 
	\sigma_R^2(p_\shortparallel)}{e^{|p_\shortparallel|} -
	\sigma_R(p_\shortparallel) \sigma_L(p_\shortparallel)} \,
\frac{|p_\shortparallel + 2 l k_\shortparallel|
	\sigma_L(p_\shortparallel + 2 l k_\shortparallel)
	\sigma_R^2(p_\shortparallel + 2 l
	k_\shortparallel)}{e^{|p_\shortparallel + 2 l k_\shortparallel|} -
	\sigma_R(p_\shortparallel + 2 l k_\shortparallel)
	\sigma_L(p_\shortparallel + 2 l k_\shortparallel)} \;,
\end{align}
and then we have the ones where each departure appears to the first order,
which are symmetrical: $\tilde{\gamma}_{LR}=\tilde{\gamma}_{RL}$:
\begin{align}\label{eq:glr}
&\tilde{\gamma}_{LR}(k_\shortparallel) \;=\; - \frac{1}{2^4 (x_L x_R)^2 l} \, \int
	\frac{d^3p_\shortparallel}{(2\pi)^3} \,
\frac{e^{\frac{1}{2}|p_\shortparallel|-\frac{1}{2}|p_\shortparallel
	+ 2 l k_\shortparallel|}}{e^{|p_\shortparallel|} -
\sigma_R(p_\shortparallel) \sigma_L(p_\shortparallel)} 
	|p_\shortparallel| \, |p_\shortparallel + 2 l k_\shortparallel| \;
\sigma_R(p_\shortparallel) \sigma_L(p_\shortparallel) \nonumber\\
& \times \sigma_R(p_\shortparallel + 2 l k_\shortparallel)
\sigma_L(p_\shortparallel + 2 l k_\shortparallel) 
\Big[1 +
\frac{\sigma_R(p_\shortparallel + 2 l k_\shortparallel)
\sigma_L(p_\shortparallel + 2 l k_\shortparallel)}{e^{|p_\shortparallel+ 2
	l k_\shortparallel|} - \sigma_R(p_\shortparallel+ 2 l k_\shortparallel)
	\sigma_L(p_\shortparallel+ 2 l k_\shortparallel)}
	\Big]\;.
	\end{align}
	
The term proportional to $\tilde\gamma_{LL}$($RR$) describes corrections to the Casimir energy produced by the inhomogeneities
of the $L$($R$) plate, while the term proportional to 
$\tilde\gamma_{LR}$ is the relevant one for the computation of 
lateral forces and torques induced by the interaction between the 
inhomogeneities on both mirrors.

The rather complex expressions can be simplified for ``quasi-Dirichlet" mirrors for which $x_{L,R}\gg 1$. In this limit
we have $\sigma_{L,R}(\rho)\simeq 1-\rho/x_{L,R}$ and therefore
the form factors $\tilde{\gamma}_{\alpha\beta}$ can be approximated
by
\begin{equation}\label{eq:gllqD}
	x_L^4\tilde{\gamma}_{LL}(k_\shortparallel) =x_R^4\tilde{\gamma}_{RR}(k_\shortparallel)= -\frac{1}{8\, l} \int
	\frac{d^3p_\shortparallel}{(2\pi)^3} \; 
	\frac{|p_\shortparallel||p_\shortparallel + 2 l k_\shortparallel|}{(e^{|p_\shortparallel|} -
	1)(1- e^{-|p_\shortparallel + 2 l k_\shortparallel|})}
	\left(1- \frac{e^{-|p_\shortparallel + 2 l k_\shortparallel|}}{2}\right)
\end{equation}
and
\begin{equation}\label{eq:glrqD}
(x_L x_R)^2\tilde{\gamma}_{LR}(k_\shortparallel) \;=\; - \frac{1}{16\,l} \, \int
	\frac{d^3p_\shortparallel}{(2\pi)^3} \,
\frac{e^{\frac{1}{2}|p_\shortparallel|+\frac{1}{2}|p_\shortparallel
	+ 2 l k_\shortparallel|}|p_\shortparallel| \, |p_\shortparallel + 2 l k_\shortparallel|}{(e^{|p_\shortparallel|} -1)(e^{|p_\shortparallel+ 2
	l k_\shortparallel|} - 1)}\, ,
\end{equation}
where we omitted a $k_\shortparallel$-independent term 
in $\tilde{\gamma}_{LL}$ and $\tilde{\gamma}_{RR}$.  We can see that in this limit the 
$\gamma_{\alpha\beta}$ are all of the same order of magnitude
if $x_R \simeq x_L$. 

It is interesting to compare these results with the one for  curved Dirichlet mirrors. For simplicity we will compare the interaction energy between Dirichlet mirrors
(a curved mirror  in front of a  planar 
one) with that of a Dirichlet mirror (R) in front of a quasi-Dirichlet planar mirror with inhomogeneities (L). The  form factor for the former configuration, that we will denote here by  $\tilde\gamma_G(k_\shortparallel)$ is similar 
to $\tilde\gamma_{LL}$  \cite{formfactor} :
\begin{equation}\label{eq:gammageo}
	\tilde\gamma_G(k_\shortparallel)= -\frac{1}{16\, l^5} \int
	\frac{d^3p_\shortparallel}{(2\pi)^3} \; 
	\frac{|p_\shortparallel||p_\shortparallel + 2 l k_\shortparallel|}{(e^{|p_\shortparallel|} -
	1)(1- e^{-|p_\shortparallel + 2 l k_\shortparallel|})}\, .
	\end{equation}
Note that the integrands of $\tilde\gamma_{LL}$ and $\tilde\gamma_G$ differ in a term proportional to $e^{-|p_\shortparallel+ 2
	l k_\shortparallel|} $, and therefore they have the same behavior in the limit $\vert k_\shortparallel\vert l\gg 1$. Indeed, one can easily 
	show that  both are proportional
	to $\vert k_\shortparallel\vert l$ in this limit. Therefore,  one can mimic a perfect mirror with non-planar geometry  with inhomogeneities on a planar mirror. The
	departure from the planar geometry of a Dirichlet mirror, $\eta_G(\mathbf x_\shortparallel)$, should be proportional to the departure from 
	the constant potential $\eta_L(\mathbf x_\shortparallel)$. This is valid when the scale of variation of the geometry (and of the inhomogeneities)
	is  much smaller that the distance between mirrors.  The opposite limit will be studied in the next section.


\section{Derivative expansion}\label{sec:de}
Let us now consider the case in which the inhomogeneities are smooth, that is, its scale of variation is much larger than $l$. 
We may extract from the results of the previous Section the would be
zero-order in the DE (`PFA')
term. Indeed, from Eq.\eqref{E(0)} we find
\begin{equation}
	E_{PFA}[V_L,V_R] \;=\;\int d^2{\mathbf x}_\shortparallel \; {\mathcal
	E}\big(l V_L({\mathbf x}_\shortparallel), l V_R({\mathbf x}_\shortparallel)\big) \;,
\end{equation}
with
\begin{equation}\label{eq:edens}
	{\mathcal E}(x_L,x_R) \,=\, \frac{1}{l^3} \, f(x_L,x_R) \;\;,\;\;\;
	f(x_L,x_R) = \frac{1}{32 \pi^2} \int_0^\infty d\rho
	\rho^2 \,\log\big(  1 - \frac{x_L}{\rho + x_L} \frac{x_R}{\rho +
	x_R} e^{-\rho} \big) \; . 
\end{equation}
In Fig. 1 we plot the function $f(x_L,\infty)$, that describes the PFA to the interaction between
a Dirichlet (R) and an inhomogeneous mirror (L). As expected, $f$ vanishes in the limit of transparent
mirrors ($x_L\to 0)$, and tends to its Dirichlet value ($-\pi^4/45$) for $x_L\gg 1$. It interpolates monotonically between these two extreme cases.

\begin{figure}[!ht]
\centering
\includegraphics[scale=0.4]{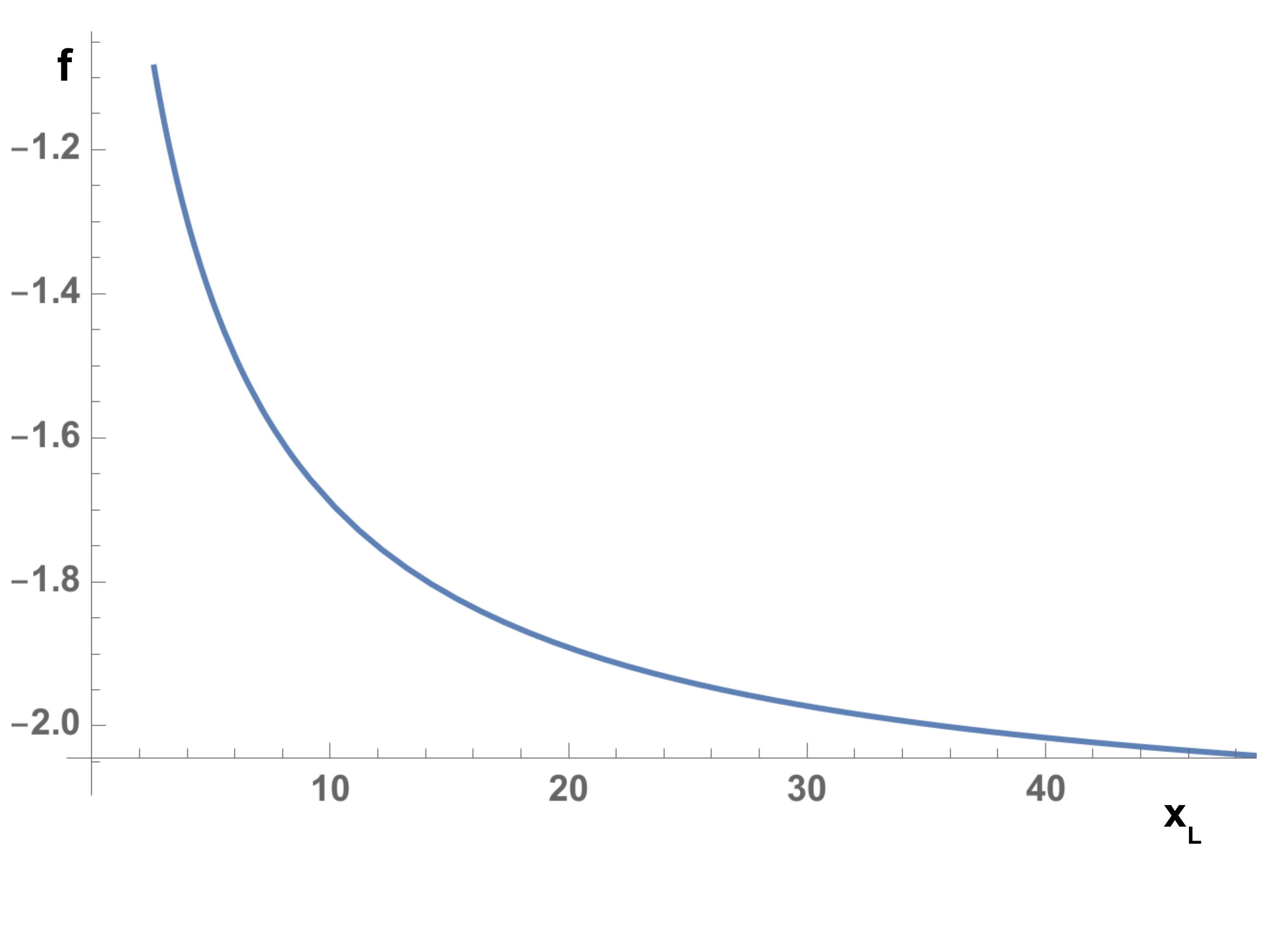} 
\caption{\label{fig1} The function $f(x_L,\infty)$ that describes the interaction between a Dirichlet mirror and an inhomogeneous
mirror. It is negative and monotonically decreasing, vanishes linearly as $x_L\to 0$, and tends to $-\pi^4/45$ as $x_L\to\infty$.}
\end{figure}

To obtain the second-order term in a DE, which we denote by $E^{(2)}_{DE}$,
we follow our previous works \cite{DE1,DE2}.
We need to extract the term which is quadratic in ${\mathbf
k}_\shortparallel$ from each $\tilde{\gamma}_{\alpha\beta}$. Namely, for
small $k_\shortparallel$:
\begin{equation}
	\tilde{\gamma}_{\alpha\beta}(k_\shortparallel) \;=\; c_{\alpha\beta}\,
	k_\shortparallel^2 \,+\, {\mathcal O}(k_\shortparallel^3) \;\;
	\;.
\end{equation}
We have found:
\begin{equation}\label{eq:crr}
	c_{RR} \,=\,\frac{1}{48\pi^2 x_R^3 v_R} \int_0^\infty d\rho \rho^2 
	\Big[ [g(p)]^2 
	\,-\,\frac{4 x_L x_R^5}{(\rho + x_L) (\rho + x_R)^4}  
	\frac{1}{(\rho + x_R)e^\rho - \frac{x_L x_R}{\rho + x_L}} \Big]\;.  
\end{equation}
where:
\begin{equation}
 g(\rho) = -\frac{x_L x_R^2 \left(e^\rho (\rho+x_R) \left(\rho^3+\rho^2 (x_L+x_R+2)+ \rho x_L (x_R+1)-x_L x_R\right)+x_L
    x_R^2\right)}{(\rho+x_R)^2 \left(x_L x_R-e^\rho (\rho+x_L) (\rho+x_R)\right)^2}
\end{equation}
$c_{LL} = c_{RR}|_{R \leftrightarrow L}$, and:
\begin{align}\label{eq:clr}
c_{LR} \,=\,\frac{l}{48\pi^2 (x_L x_R)^2} \int_0^\infty d\rho \rho^2 
	\left\{ \partial_\rho \Big[ \frac{e^{\frac{\rho}{2}} \, \rho
\, \sigma_L(\rho) \sigma_R(\rho)}{e^\rho  - \sigma_L(\rho)
	\sigma_R(\rho)}\Big]  \right\}^2 \;.
\end{align}
Therefore, going back to configuration space, the correction to the PFA reads
\begin{equation}
E_{DE}^{(2)}=-\frac{1}{2}\sum_{\alpha,\beta} \int d^2{\mathbf
x}_\shortparallel c_{\alpha\beta}(lV_L,lV_R)\nabla V_\alpha\cdot\nabla V_\beta\, .
\end{equation}

It is again instructive to analyze the case of quasi-Dirichlet mirrors. As before, we will consider the case in which the 
R mirror is perfect ($x_R\to \infty$), and the L mirror is quasi-perfect.  Expanding the leading order result for $x_L\gg 1$ we obtain
\begin{equation}
E_{PFA}=-\frac{A\, \pi^2}{1440\,  l^3}\left (1-\frac{3}{l}\langle\frac{1}{V_L}\rangle\right )\, ,
\end{equation}
where $\langle ...\rangle$ denotes the mean value over the surface. On the other hand, in this limit
\begin{equation}
c_{LL}\simeq\frac{k} {48\pi^2 l^3V_L^4} \, ,
\end{equation}
with
\begin{equation}
k=\int_0^\infty d\rho\rho^2 \left[\left(\frac{1+e^\rho(\rho-1)}{(e^\rho-1)^2}\right)^2-\frac{4}{e^\rho-1}\right]\simeq -9.14
\end{equation}
As a consequence,   the NTLO correction reads
\begin{equation}
E_{DE}^{(2)}\simeq  \frac{9.14}{96\pi^2 l^3}\int d^2{\mathbf x}_\shortparallel  \frac{\left(\nabla V_L\right)^2}{V_L^4}\, .
\end{equation}
These results can be written in a simpler way by introducing a sort of ``conductivity" $\rho_L=V_L^{-1}$ (note that $\rho_L\to 0$
for perfect mirrors)
\begin{equation}
\frac{E_{DE}}{A}=-\frac{ \pi^2}{1440\,  l^3}\left (1-\frac{3}{l}\langle\rho_L\rangle\right ) +\frac{9.14}{96\pi^2 l^3} \langle \left(\nabla \rho_L\right)^2
\rangle\, .
\end{equation}

\section{Conclusions}\label{sec:conc}
In this paper we evaluated the vacuum energy for a scalar field in the
presence of two inhomogeneous thin plates. The interaction between the
field and the plates is modelled by potentials localized on the plates. The
calculation generalizes previous works on $\delta$-potentials and can be
considered as a toy model for the interaction of the electromagnetic field
with microstructured flat mirrors.

From a mathematical point of view, we used the GY theorem to compute the corresponding functional determinant, and arrived to a formal result for the vacuum energy, Eq.\eqref{eq:lifshitz},  that can be thought as  a generalized Lifshitz formula, in which the reflection coefficients are replaced by nonlocal operators that become local for homogeneous plates.

We have obtained explicit integral expressions for the vacuum energy in two different situations that involve either small or smooth inhomogeneities. For small inhomogeneities, we compared the result with the corresponding one for Dirichlet plates with non trivial geometries, pointing out that
when the scales of variation of the geometry and the inhomogeneity are much smaller than the distance between mirrors,  geometry can traded off by inhomogeneity.   For smooth inhomogeneities, we used the DE as a expansion to obtain the NTLO correction to the PFA approximation. 

The formalism developed in this paper can be adapted to compute the vacuum energy in the electromagnetic case, including an inhomogeneous medium between plates. The calculation of Casimir-Polder forces on atoms near an inhomogeneous plate is also of interest. We plan to address these issues in a forthcoming work.

\section*{Acknowledgments}
This research was supported by ANPCyT, CONICET, and UNCuyo. 

\end{document}